\documentclass[
    ,final            % use final for the camera ready runs
  ]
  {aipproc}

\layoutstyle{8x11double}
\begin{document}

\title{Time Averaged VHE Spectrum of Mrk 421 in 2005.}

\classification{
%%<Replace this text with PACS numbers; choose from this list:
%%\texttt{http://www.aip.org/pacs/index.html}>
%95.55.-n	Astronomical and space-research instrumentation (see also 94.80.+g Instrumentation for space plasma physics, ionosphere, and magnetosphere)
%95.55.Ka	X- and ?-ray telescopes and instrumentation
%95.75.-z	Observation and data reduction techniques; computer modeling and simulation
%95.75.Fg	Spectroscopy and spectrophotometry
%95.85.-e	Astronomical observations (additional primary heading(s) must be chosen with these entries to represent the astronomical objects and/or properties studied)
%95.85.Pw	?-ray
%98.54.-h	Quasars; active or peculiar galaxies, objects, and systems
%98.54.Cm	Active and peculiar galaxies and related systems (including BL Lacertae objects, blazars, Seyfert galaxies, Markarian galaxies, and active galactic nuclei)
95.55.Ka, 95.75.Fg, 95.85.Pw, 98.54.Cm
}


\author{M.~.K.~Daniel}{
  address={School of Physics and Astronomy, University of Leeds, Leeds, LS2 9JT. U.K.}
  ,altaddress={now at Department of Physics, University of Durham, South Road, Durham, DH1 3LE. U.K.} % additional visiting address
}

\author{the VERITAS Collaboration}{
  address={see \url{http://veritas.sao.arizona.edu/} for a list of members}
}

\begin{abstract}
The blazar Mrk421 was observed independently, but contemporaneously, in 2005 at 
TeV energies by MAGIC, the Whipple 10m telescope, and by a single VERITAS 
telescope during the construction phase of operations. A comparison of the time 
averaged spectra, in what was a relatively quiescent state, demonstrates the 
level of agreement between instruments. In addition, the increased sensitivity
of the new generation instruments, and ever decreasing energy thresholds, 
questions how best to compare new observational data with archival results.
\end{abstract}

\maketitle

%%%%%%%%%%%%%%%%%%%%%%%%%%%%%%%%%%%%%%%%%%%%
%% MAINMATTER
%%%%%%%%%%%%%%%%%%%%%%%%%%%%%%%%%%%%%%%%%%%%

\section{Introduction}
\begin{figure}[ht]
% places the figures side by side - we want them in a column
  \includegraphics[width=0.48\textwidth]{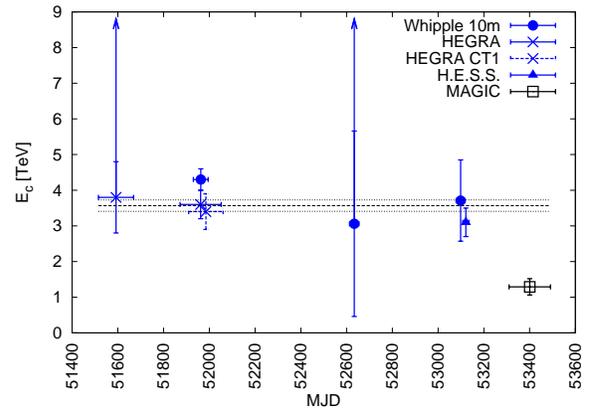}
  \caption{Values for the fit of a power law with
          exponential cut-off term to the Mrk 421
          spectrum available in the literature \cite{Krennrich2001,
          Krennrich2002, Aharonian2002, Aharonian2003, Rebillot2006,
          Daniel2005, Aharonian2005, Mazin2007}. 
          Plotted as a function of epoch.}
\label{fig:CutoffByTime}
\end{figure}

\begin{figure}
% places the figures side by side - we want them in a column
%  \includegraphics[height=.3\textheight]{Fig01.eps}
  \includegraphics[width=.48\textwidth]{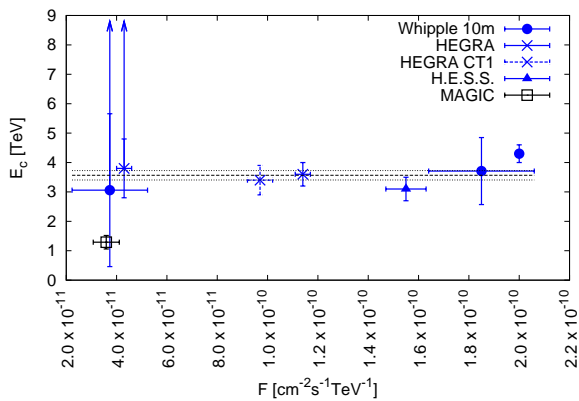}
%  \includegraphics[width=.5\textwidth]{Fig03.eps}
%  \caption{Values for the fit of a power law with
%          exponential cut-off term to the Mrk 421
%          spectrum available in the literature \cite{Krennrich2001,
%          Krennrich2002, Aharonian2002, Aharonian2003, Rebillot2006,
%          Daniel2005, Aharonian2005, Mazin2007}.}
   \caption{as figure~\ref{fig:CutoffByTime}, 
           but the cut-off energy is plotted as a function of flux level.}
\label{fig:CutoffByFlux}
\end{figure}

\begin{figure}
% places the figures side by side - we want them in a column
%  \includegraphics[height=.3\textheight]{Fig01.eps}
%  \includegraphics[width=.5\textwidth]{Fig02.eps}
  \includegraphics[width=.48\textwidth]{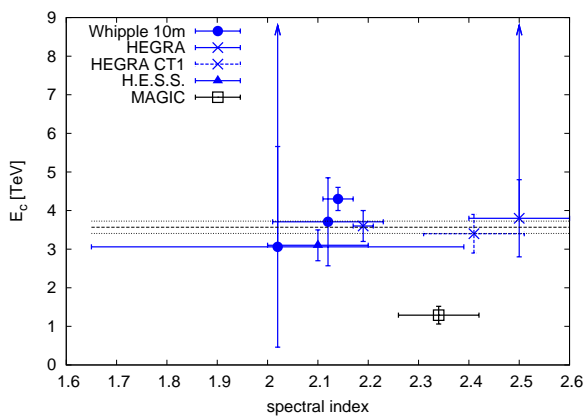}
%  \caption{Values for the fit of a power law with
%          exponential cut-off term to the Mrk 421
%          spectrum available in the literature \cite{Krennrich2001,
%          Krennrich2002, Aharonian2002, Aharonian2003, Rebillot2006,
%          Daniel2005, Aharonian2005, Mazin2007}.}
   \caption{as figure~\ref{fig:CutoffByTime}, 
           but the cut-off energy is plotted as a function of the derived 
           spectral index.}
\label{fig:CutoffByIndex}
\end{figure}

%Mrk 421 and its history?
%the experiments
Since its discovery as the first extragalactic source of very high energy (VHE,
$E>100\,$GeV) gamma rays, Mrk 421 has been intensively studied. The spectral
energy distribution (SED) shows the two characteristic non-thermal peaks of a
high energy peaked BL-Lac (HBL) system. 
The VHE end of the spectrum is complicated by attenuation by the extragalactic 
background light (EBL) which will add a characteristic absorption to the
spectrum \cite{Dwek05}. Spectral variability as a function of flux level has 
been observed, the spectrum hardening with increasing flux level 
\cite{Krennrich2002, Aharonian2002}, and curvature being readily 
apparant in time averaged spectra at the highest flux levels 
\cite{Krennrich2001, Daniel2005, Aharonian2005}. The most favoured spectral fit
form for the curvature has been that of a power-law with an exponential cut-off 
\begin{equation}
%$
\frac{\mathrm{d}N}{\mathrm{d}E} = 
 F_{c}
 \exp{\left(-\frac{E}{E_{c}}\right)}
 E^{-\alpha}
%\label{eqn:cutoff}
%$
\end{equation}
where $F_{c}$ is the flux at 1\,TeV, $E_{c}$ is the characteristic cut-off
energy in TeV and $\alpha$ is the spectral index. 
In spite of the spectral variability, the value for the cut-off has seemed a 
relatively consistent feature between experiments and epochs (see 
figures \ref{fig:CutoffByTime}, \ref{fig:CutoffByFlux}, \ref{fig:CutoffByIndex}) at an average value of $E_C = 3.57\pm0.16$\,TeV. 
At low/intermediate flux levels there had not been sufficient improvement in a
$\chi^2$ fit to favour a curved spectrum over that of a pure power-law. That was 
until measurements made by the new generation instrument MAGIC during 2005 
showed compelling evidence in the time averaged spectrum favouring a curved 
spectral shape \cite{Mazin2007}, although with a much lower value for a cut-off 
than had been seen previously. Given that a change in the cut-off would 
demonstrate spectral curvature to unequivocally be intrinsic to the source 
rather than just attenuation due to the EBL it is important to confirm these 
findings. 
The Whipple 10\,m and the first of the VERITAS telescopes also observed 
Mrk\,421 during 2005 and this paper compares the results of these three 
independent sets of measurements of the time averaged spectrum.

\section{Observations and Data Analysis}
\begin{figure}
  \includegraphics[width=\textwidth]{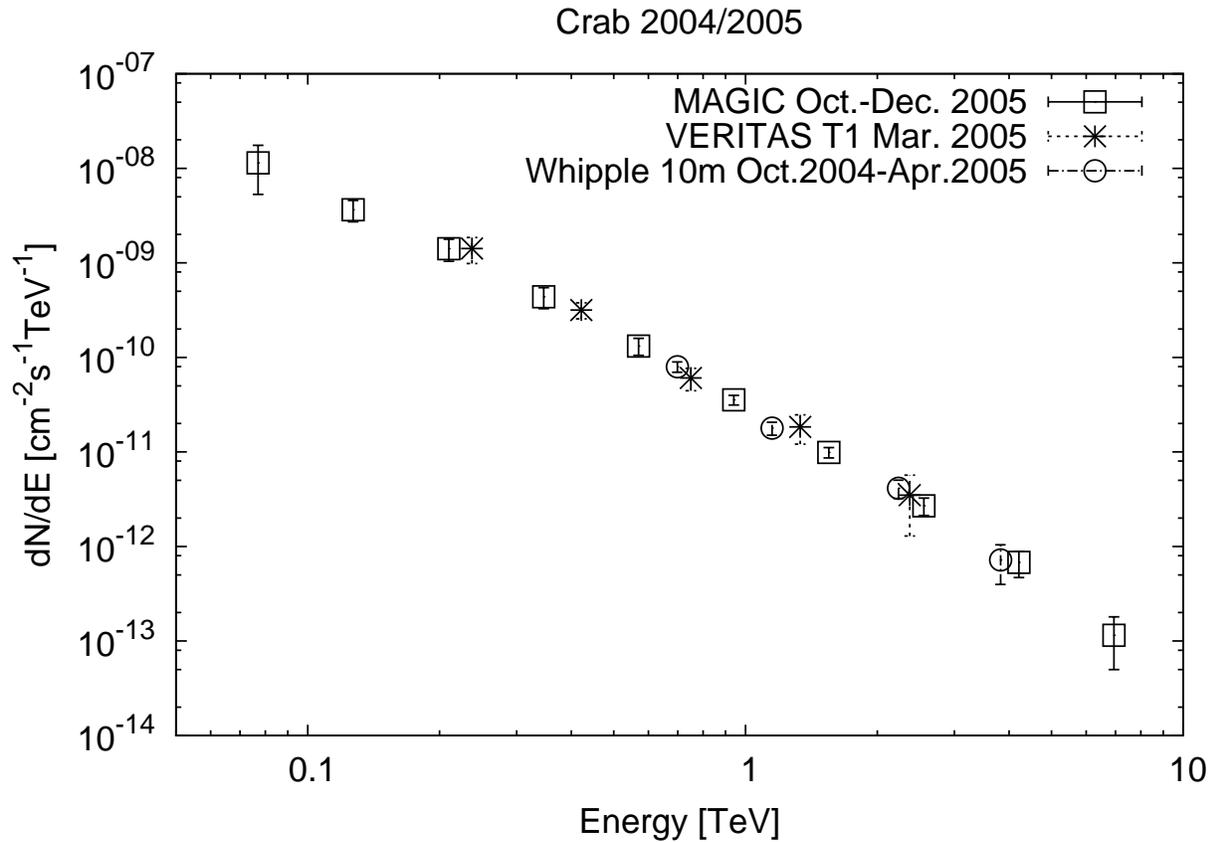}
  \caption{Comparison of the spectral reconstruction for the three different 
          telescopes for the standard candle of the Crab nebula.
         }
  \label{fig:Crab}
\end{figure}

The Whipple 10\,m is described in \cite{Kildea07}. 
The VERITAS T1 system is described in \cite{Holder06}, even though the array was 
in the construction phase during this time frame this dataset was taken when T1 
had begun taking regular observations. The MAGIC telescope is described in 
\cite{Ferenc05}. 
As a crosscheck between the different experiments and their analysis procedures
the reconstruction of the Crab spectrum for the equivalent epoch was compared,
as shown in figure~\ref{fig:Crab}, and demonstrates consistent results.

\begin{figure}[h]
% places the figures side by side
%  \includegraphics[width=.5\textwidth]{Fig04.eps}
  \includegraphics[width=\textwidth]{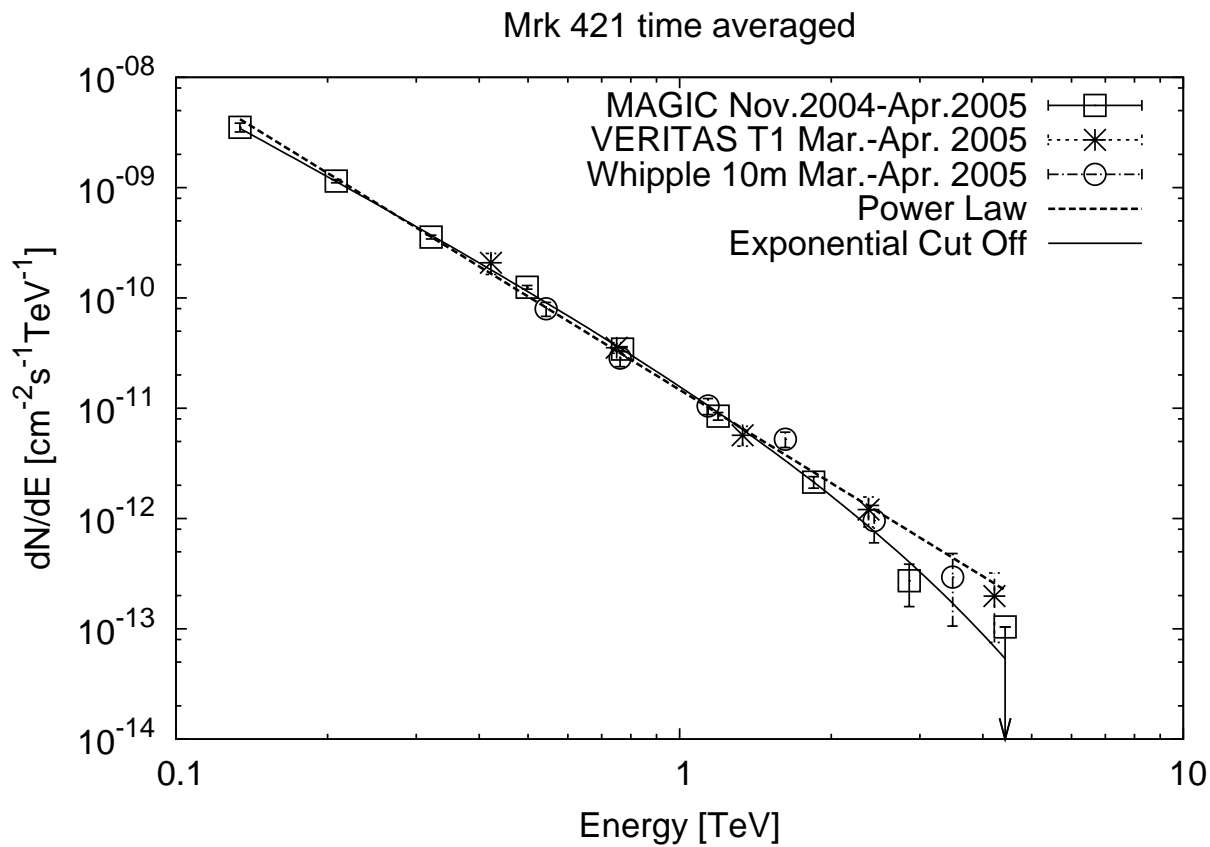}
%  \caption{Comparison of the spectral reconstruction for the three different 
%          telescopes. On the left is the standard candle of the Crab nebula, on
%          the right for the Mrk\,421 observations.}
  \caption{Comparison of the spectral reconstruction for the three different 
          telescopes for the time averaged Mrk\,421 observations. 
          The lines are the best fits to the combined data sets.
         }
  \label{fig:Spectra}
\end{figure}

The MAGIC observations of Mrk\,421 are detailed in \cite{Mazin2007} and consist 
of 26 hours of on-source time taken between November 2004 and April 2005. 
The Whipple 10\,m Mrk\,421 observations consist of 10 hours on-source data taken 
during March-April of 2005 as part of the regular blazar monitoring program. 
The data were reduced and spectrum derived according to the method described in 
\cite{Grube2007}. 
The VERITAS T1 data consist of 8 hours of on source observations also taken 
during March and April 2005. It can be seen from the tables that whilst the 
MAGIC observations overwhelmingly favour a curved spectrum the fits to the 
other two show no improvement to those for a pure power-law. 
The time averaged Mrk\,421 spectrum for the three experiments are shown together
in figure~\ref{fig:Spectra}. 
The fits to a pure power-law function are summarised in 
table~\ref{tab:power} and to a power-law with an exponential cut-off 
are summarised in table~\ref{tab:cutoff} respectively. 
A fit to all of the flux points
combined still favours a curved spectrum, but with a cut-off 
$E_c = 1.65 \pm 0.31$\,TeV which is still much lower than previously measured 
in the Mrk\,421 spectrum.

\section{Discussion \& Conclusions}
Even though the flux points between the different experiments agree well, from a
visual perspective at any rate, they have different conclusions on the best fit 
of the spectral form: with the MAGIC data clearly favouring a curvature of the 
low flux level spectrum that would appear different with previous results at
higher flux levels. 
Whilst the data are all time averaged for the same epoch they do cover slightly 
different time periods and are not truly simultaneous; if the cut-off in the
Mrk\,421 spectrum represent intrinsic changes in the emission region then the
different experiments could see different behaviour if sampling at different
times. The consistency of the previous generation instruments results does not 
sit well with this conclusion even if the cut-off varies as a function of flux 
level. 
The obvious difference between these different single dish systems is the lower
threshold energy of MAGIC due to a much larger mirror area. When the MAGIC
points corresponding only to the energy range of the smaller instruments, and
the previous generation of instruments, are used then a curved spectrum is still
favoured, both individually and when combined with all flux points. The value
for this cut-off is $E_c = 3.24\pm2.13$\,TeV, now in agreement with the 
historical and present data. The larger error bars down to a smaller number of 
points over a much reduced energy range that are less constraining to the fit. 
This clearly demonstrates that the spectrum of Mrk\,421 is not merely cut-off at 
the highest energies, but exhibits curvature throughout the VHE spectrum. It may 
be prudent to investigate other fit functions (such as a broken power law, or a
log-parabolic form found to improve fits in X-ray spectrum fits)
to describe the spectral curvature with the new generation of lower threshold 
instruments, and particularly in light of the 5 orders of magnitude in energy 
range that will be available in joint multiwavelength observations with GLAST.

%%%%%%%%%%%%%%%%%%%%%%%%%%%%%%%%%%%%%%%%%%%%%%%%
%% BACKMATTER
%%%%%%%%%%%%%%%%%%%%%%%%%%%%%%%%%%%%%%%%%%%%%%%%

\begin{theacknowledgments}
The author wishes to to thank Daniel Mazin for providing the MAGIC data and
helpful comments. 
This research is supported by grants from the U.S. Department of
Energy, the U.S. National Science Foundation, and the Smithsonian
Institution, by NSERC in Canada, by PPARC in the UK and by Science
Foundation Ireland.
\end{theacknowledgments}

%%%%%%%%%%%%%%%%%%%%%%%%%%%%%%%%%%%%%%%%%%%%%%%%
%% The bibliography can be prepared using the BibTeX program or
%% manually.
%%
%% The code below assumes that BibTeX is used.  If the bibliography is
%% produced without BibTeX comment out the following lines and see the
%% aipguide.pdf for further information.
%%
%% For your convenience a manually coded example is appended
%% after the \end{document}
%%%%%%%%%%%%%%%%%%%%%%%%%%%%%%%%%%%%%%%%%%%%%%%%

%%%%%%%%%%%%%%%%%%%%%%%%%%%%%%%%%%%%%%%%%%%%%%%%
%% You may have to change the BibTeX style below, depending on your
%% setup or preferences.
%%
%%
%% For The AIP proceedings layouts use either
%%%%%%%%%%%%%%%%%%%%%%%%%%%%%%%%%%%%%%%%%%%%

\bibliographystyle{aipproc}   % if natbib is available
%\bibliographystyle{aipprocl} % if natbib is missing

%%%%%%%%%%%%%%%%%%%%%%%%%%%%%%%%%%%%%%%%%%%
%% You probably want to use your own bibtex database here
%%%%%%%%%%%%%%%%%%%%%%%%%%%%%%%%%%%%%%%%%%%
\bibliography{Gamma08_mkd_Mrk421_2005}

%%%%%%%%%%%%%%%%%%%%%%%%%%%%%%%%%%%%%%%%%%%
%% Just a reminder that you may have to run bibtex
%% All of it up to \end{document} can be removed
%% if you don't like the warning.
%%%%%%%%%%%%%%%%%%%%%%%%%%%%%%%%%%%%%%%%%%%
\IfFileExists{\jobname.bbl}{}
 {\typeout{}
  \typeout{******************************************}
  \typeout{** Please run "bibtex \jobname" to optain}
  \typeout{** the bibliography and then re-run LaTeX}
  \typeout{** twice to fix the references!}
  \typeout{******************************************}
  \typeout{}
 }

\begin{table}[h]
\begin{tabular}{lccc}
\hline
    \tablehead{1}{l}{b}{Experiment} % \\ for newline
  & \tablehead{1}{c}{b}{$F_{c}$
                      $[\mathrm{cm}^{-2}\mathrm{s}^{-1}]$}
  & \tablehead{1}{c}{b}{$\alpha$}   
  & \tablehead{1}{c}{b}{$\chi^{2}(n)$} \\
\hline
MAGIC & $ \left( 1.56 \pm 0.05 \right) \times 10^{-11} $ 
      & $2.80 \pm 0.02$ 
      & $53.9(6)$ \\
10m   & $ \left( 2.04 \pm 0.16 \right) \times 10^{-11} $ 
      & $2.80 \pm 0.17$
      & $0.74(4)$ \\
T1    & $ \left( 1.17 \pm 0.10 \right) \times 10^{-11} $ 
      & $3.04\pm0.18$
      & $0.31(4)$\\
combined%\tablenote{normalised according to flux at 1\,TeV} 
      & $ \left( 1.47 \pm 0.04 \right) \times 10^{-11} $ 
      & $2.81 \pm 0.02$
      & $60.69(18)$\\
\hline
\end{tabular}
\caption{Fits to a pure power-law spectrum.}
\label{tab:power}
\end{table}

\begin{table}
\begin{tabular}{lcccc}
\hline
    \tablehead{1}{l}{b}{Experiment} % \\ for newline
  & \tablehead{1}{c}{b}{$F_{c}$
                      $[\mathrm{cm}^{-2}\mathrm{s}^{-1}]$}
  & \tablehead{1}{c}{b}{$E_{c}$ $[\mathrm{TeV}]$}
  & \tablehead{1}{c}{b}{$\alpha$}   
  & \tablehead{1}{c}{b}{$\chi^{2}(n)$} \\
\hline
MAGIC & $ \left( 3.59 \pm 0.05 \right) \times 10^{-11} $ 
      & $1.28 \pm 0.23$ 
      & $2.3\pm0.08$ 
      & $5.89(5)$ \\
10m   & $ \left( 2.27 \pm 0.52 \right) \times 10^{-11} $ 
      & $10.64 \pm 59.18$ 
      & $2.69 \pm 0.62$
      & $0.73(3)$ \\
T1    & $ \left( 1.17 \pm 0.10 \right) \times 10^{-11} $ 
      & $\infty$
      & $3.04\pm0.18$
      & $0.31(3)$\\
combined%\tablenote{normalised accroding to flux at 1\,TeV} 
      & $ \left( 2.88 \pm 0.34 \right) \times 10^{-11} $ 
      & $1.65 \pm 0.31$
      & $2.42 \pm 0.07$
      & $19.46(17)$\\
\hline
\end{tabular}
\caption{Fits to a powerlaw spectrum with an exponential cutoff.}
\label{tab:cutoff}
\end{table}

\end{document}